\documentclass[a4paper,11pt]{article}
\pdfoutput=1
\usepackage{jheppub}
\usepackage{amsmath,amssymb,latexsym}
\usepackage{braket}
\usepackage{bm}
\usepackage{hepunits}
\usepackage[svgnames]{xcolor}

\newcommand{\nn}{\nonumber}

\newcommand{\mG}{\mathcal{G}}
\newcommand{\aP}{\left\langle P \right\rangle}

\allowdisplaybreaks

\def\eref#1{(\ref{#1})}
\newcommand{\bea}{\begin{eqnarray}}
	\newcommand{\eea}{\end{eqnarray}}
\newcommand{\bean}{\begin{eqnarray*}}
	\newcommand{\eean}{\end{eqnarray*}}

\title{
Module Intersection and  Uniform  Formula for Iterative Reduction of One-loop Integrals
}

\author[a]{Jiaqi Chen,}
\author[a,b,c,d]{Bo Feng}

\affiliation[a]{Beijing Computational Science Research Center, Beijing 100084, China}
\affiliation[b]{Zhejiang Institute of Modern Physics, Zhejiang University, Hangzhou, 310027, P. R. China}
\affiliation[c]{ Center of Mathematical Science, Zhejiang University, Hangzhou, 310027, P. R. China}
\affiliation[d]{Peng Huanwu Center for Fundamental Theory, Hefei, Anhui 230026, China}

\emailAdd{jiaqichen@csrc.ac.cn}
\emailAdd{fengbo@csrc.ac.cn}

\abstract{In this paper, we develop an iterative sector-level reduction strategy for
Feynman integrals, which bases on  module intersection in the Baikov representation and auxiliary vector for tensor structure. Using this strategy
we have studied the reduction of general one-loop integrals, i.e., integrals having arbitrary  tensor structures and arbitrary power for propagators.
Inspired by these studies, a uniform and compact formula that iteratively reduces all one-loop integrals has been written down, where messy polynomials in integration-by-parts (IBP) relations have organized themselves to Gram determinants.}

\begin{document}

\maketitle


\section{Introduction}
Particle physics is highly developed and comes to the era of precise measurement, which calls for high precision theoretical predictions. One of the most important parts of theoretical predictions is perturbation computations in quantum field theory. For such computations,
reducing   general Feynman integrals into master integrals plays an important role \cite{Chetyrkin:1981qh}. It not only greatly reduces the number of integrals to be calculated, but also produces the differential equations \cite{Kotikov:1990kg,Gehrmann:1999as,Argeri:2007up,Henn:2014qga} for solving the master integrals. Based on  differential equations,
many analytical or numerical methods and packages for solving master integrals have been developed
 \cite{Henn:2013pwa,Moriello:2019yhu,Bonciani:2019jyb,Frellesvig:2019byn,Hidding:2020ytt,Liu:2017jxz,Liu:2022tji,Liu:2022mfb,Liu:2022chg,Liu:2021wks,Liu:2020kpc,Armadillo:2022ugh}, which can be widely applied to multi-loop and multi-scale processes.

The reduction problem can be roughly divided into the reduction of loop momenta in the numerator (called "tensor reduction" or TR) and the reduction of propagators in the denominator with general powers (called "denominator reduction" or DR). These two kinds of reductions (i.e., TR and DR) are tightly connected\footnote{For example, in \cite{Feng:2022uqp} one has used the tensor reduction to solve reduction with  general denominators for one-loop integrals.}. The well-established  method to do reduction is the integration-by-parts (IBP) method introduced in \cite{Chetyrkin:1981qh}. When  combining with the Laporta algorithm \cite{Laporta:2000dsw},  several widely used packages have been developed such as \cite{Smirnov:2019qkx,vonManteuffel:2012np,Lee:2012cn,Klappert:2020nbg}. IBP relations give many linear relations between various scalar integrals and  solving them gives  relations between a given integral and the master integrals (i.e., the reduction). When it comes to multi-loop and multi-scale integrals required by higher precision theoretical predictionw, the size of linear equations generated by IBP relations increases dramatically. Solving them
becomes more and more difficult.
Motivated by this, many developments of reduction methods have been explored for higher loops\footnote{For one-loop integrals, two very efficient methods are the unitarity cut method \cite{Bern:1994zx,Bern:1994cg,Britto:2004nc,Britto:2005ha} for reduction at the integral level and OPP method \cite{Ossola:2006us} for reduction at the integrand level. } \cite{Gluza:2010ws,Peraro:2019svx,Chestnov:2022alh,Mastrolia:2018uzb,Frellesvig:2019kgj,
Frellesvig:2019uqt,Weinzierl:2020xyy,Mizera:2019ose,Frellesvig:2020qot,
Liu:2018dmc,Guan:2019bcx,Larsen:2015ped,Larsen:2016tdk,Zhang:2016kfo,
Georgoudis:2016wff,Georgoudis:2017iza,Bohm:2017qme,Bohm:2018bdy,Bendle:2019csk,
Boehm:2020zig,Bendle:2021ueg,Feng:2022iuc}.

Most of the above methods solve a hugely redundant linear system with many equations using programs. On one side, it makes the computation automatically. On the other side, it works likes in a black box. Although we have obtained the wanted results, we lost a full understanding of these equations. Knowing these details not only helps overcome difficulties faced by these methods but also gives a better understanding of the mathematical structures of general Feynman integrals.

	For example, there are some iterative structures in reduction. Let us consider   the bubble topology with propagators:
	\begin{align}
	z_1=l^2-m_1^2,\ \ \  z_2=(l-p)^2-m_2^2.~~~\label{1.1}
	\end{align}
There are three master integrals defined by
\bea \vec{f}=\{\text{I}_{1,0},\text{I}_{0,1},\text{I}_{1,1}\},~~~~\text{I}_{a_1,a_2} \equiv  \int\frac{d^d l}{i(\pi)^{d/2}}\frac{1}{z_1^{a_1} z_2^{a_2}}~.~~~\label{1.2}\eea
One kind of differential equations of master integrals is given by
\bea \partial_{m_1}\vec{f}=A_{m_1}\vec{f}~.~~\label{1.3}\eea
It is easy to see when taking $\partial_{m_1}$ on $\vec{f}$ iteratively we have
\bea & &\partial_{m_1}^n\vec{f}=\{ (n-1)!\  \text{I}_{n+1,0},\ 0,\  (n-1)!\ \text{I}_{n+1,1}\}=
A_{m_1}^{(n)}\vec{f} ~~~\label{1.4}\eea
where
\bea A_{m_1}^{(n)} = \partial_{m_1}A_{m_1}^{(n-1)} + A_{m_1}^{(n-1)}.\ A_{m_1} ~~~\label{1.5}\eea
The iterative structure of $A_{m_1}^{(n)}$ in \eref{1.5} gives a very simple way to find the reduction  of $\text{I}_{n+1,0}$ and $\text{I}_{n+1,1}$ (DR type) using \eref{1.4}. This is an iterative reduction relation at topology-level\footnote{When we say a reduction is at the topology-level, we mean that the integral is written as the combination of master integrals of the same topology and master integrals of the sub-topologies. }. While the iterative relation at the topology-level is powerful, it is also not easy to get. As in this example, the $A_{m_1}$ is usually calculated by the traditional IBP method, which is difficult to get for more complicated cases.

Another well-known phenomenon of reduction is that the reduction can be organized into different sectors. Usually, the reduction in the top sector can be done easily using some tricks, for example, the maximum cut for the one-loop box in \cite{Britto:2004nc}. However, such a trick lost the information of sub-sectors. Thus if there is a way such that the reduction to its top sector is not much harder than the maximal cut, but the complete information of sub-sectors has been kept, we can carry out the whole reduction top-down (or triangulated) by treating each sub-sector as another "top sector". This motivates us to consider the iterative structure at the sector-level but not the topology-level\footnote{When we say the reduction is at the sector-level, we mean that the integral is written as the combination of master integrals of the same topology and integrals (not need to be master integrals) of  the sub-topologies. }.

In this paper, we will explore more information hidden inside the IBP relations.
The method we will introduce mainly bases on syzygy and module intersection \cite{Gluza:2010ws,Schabinger:2011dz,Larsen:2015ped,Larsen:2016tdk,Zhang:2016kfo,Georgoudis:2016wff,Georgoudis:2017iza,Bohm:2017qme,Bohm:2018bdy,Bendle:2019csk,Boehm:2020zig,Bendle:2021ueg} in Baikov representation \cite{Baikov:1996iu}.  We will give a quick review in Sec.\ref{sec2}
for Baikov representation and module intersection. The main point of our method comparing to the  module intersection in \cite{Bendle:2019csk} is that we will not generate all IBP relations with   basis obtained by the module intersection and solve these redundant linear equations. On the contrary, we pick some elements in the module intersection. These elements generate iterative reduction relations for corresponding sector. Using just a few relations, one can iteratively reduce any integral in this sector to master integrals of this sector, and keep the complete information of sub-sectors.
  In such a strategy of \textbf{iteration at the sector-level}, iterative reduction relations are easier to obtain and they avoid the huge redundancy in linear  system of traditional IBP methods. To demonstrate our method, we will present one TR example and one DR example in Sec.\ref{sec:example} and discuss the similarity of these two examples. Inspired by examples in
  the previous section, we construct an uniform formula in Sec.\ref{sec:fomula}. It solves both  TR and DR for general  one-loop integrals.
  In Sec.\ref{section-6}, we give a brief discussion for
  degenerated situations, i.e., kinematics and masses take some specific values, such as null momenta or on-shell momenta. We will find that  naively applying the uniform formula laid out in Sec.\ref{sec:fomula} sometimes does not lead to the simplest iterative relation and more careful treatment is needed. However the method introduced in Sec.\ref{section-3}
  still works well by giving the simplest iterative relation. Finally, we give a brief discussion in the Sec.\ref{sec:conc}.

\section{Baikov representation and module intersection} \label{sec2}

In this section, we will review the Baikov representation of integrals \cite{Baikov:1996iu}. In this frame, it is easier to implement the module intersection as will be discussed shortly. The
Baikov representation transforms  integrals in the standard form obtained from Feynman rules by changing integral variables from $\prod_{i} d^dl_i$ to $\prod_{j} dz_j$, where each $z_i$ represents a propagator (or related Lorentz invariant scalar product  involving loop momenta). For one-loop integrals, which are the focus of our current paper, we denote propagators and integrals as
\begin{align}
	&z_1=l^2-m_1^2,\ \ \  z_2=(l+p_1)^2-m_2^2,\ \ \  z_3=(l+p_1+p_2)^2-m_3^2, \cdots\ \nn\\
	&z_{n}=(l+p_1+\dots+p_E)^2-m_{n}^2.\nn\\
	&\text{I}_{\{a_i\}}\equiv \text{I}_{a_1,a_2,\cdots,a_{n}} \equiv  \int\frac{d^d l}{i(\pi)^{d/2}}\frac{1}{\prod_{i=1}^{n} z_i^{a_i}},~~~\label{2.1}
\end{align}
where E is the number of independent external momenta and $n=E+1$. The Baikov representation of integrals is
\begin{align}
\text{I}_{a_1,a_2,\cdots,a_{E+1}} &=  \int C_{n}(d)  \mathcal{K}^{-(d-n)/2}    \mathcal{G}(\bm{z})^{(d-n-1)/2}\frac{dz_i}{\prod_{i=1}^{n}  z_i^{a_i}}~~~\label{2.2}
\end{align}
where the constant-coefficient $C_{n}(d)$ and the Gram determinant $\mathcal{K}$ of external momenta do not involve $z_i$, so they do not affect our later discussions and can be ignored. The $\mathcal{G}(\bm{z})$ is another Gram determinant depending on both loop momentum and external momenta, i.e.,
\begin{align}
\mathcal{G}(\bm{z})=G(l,p_1,\cdots,p_E)~~~\label{2.3}
\end{align}
with $G$ defined as
\begin{equation}
G(q_1,\ldots,q_n) \equiv \det (q_i \cdot q_j) \equiv \det
\begin{pmatrix}
q_1 \cdot q_1 & q_1 \cdot q_2 & \cdots & q_1 \cdot q_n
\\
q_2 \cdot q_1 & q_2 \cdot q_2 & & \vdots
\\
\vdots & & \ddots & \vdots
\\
q_n \cdot q_1 & \cdots & \cdots & q_n \cdot q_n
\end{pmatrix}
\, .~~~\label{2.4}
\end{equation}

The well-established IBP relations can also be easily implemented in the Baikov representation.
However, the differentiation on $\mathcal{G}$ will change its power, which is equivalent to shifting the space-time dimension $d$ to different values.

To avoid such a situation, the syzygy module is introduced \cite{Larsen:2015ped}. Let us  consider the
IBP relation
\begin{align}
C \int \sum_{i=1}^n \left[ \partial_{z_i} \left( P_i   \frac{1}{\prod_{i=1}^n z_i^{a_i}} \mathcal{G}(\bm{z})^{(d-n-1)/2} \right)\right]  \prod_{i=1}^n dz_i  \label{eq:baikovibp1}
\end{align}
with $P_j$s being  polynomials of $z_i$, one can see that if these $P_i$'s are properly chosen, i.e., they satisfy
\begin{align}
\sum_i^n  (P_i  \partial_{z_i} \mG ) + P_0 \mG=0  , \label{eq:gramsyz}
\end{align}
the power of $\mathcal{G}$ will not be shifted. The relation \eref{eq:gramsyz} is a syzygy equation for the set of $(n+1)$ polynomials
\begin{align}
\left\langle \partial_{z_1} \mG ,\cdots, \partial_{z_n} \mG ,\mG \right\rangle~~~
\label{set-poly}
\end{align}
All solutions of \eref{eq:gramsyz} give the syzygy module of the set \eref{set-poly}. Putting every solution back to \eref{eq:baikovibp1} we get an IBP relation   with a given $\{a_i\}$ set,
which does not involve dimension shift.

For later convenience we define the following notations: 
\begin{align}
&\left\langle P \right\rangle  = \left\langle P_1,P_2,\cdots,P_n,P_0 \right\rangle \nn\\
&  D_{\left\langle P \right\rangle} \equiv \left\{D_{P_1},\cdots, D_{P_n} , D_{P_0}\right\} \equiv \left\{   \partial_{z_1} \left(  P_1 \bm{\cdot} \right) , \cdots, \partial_{z_n} \left(  P_n \bm{\cdot} \right) ,  \frac{d-n-1}{2} P_0 \bm{\cdot}  \right\}  \nn\\
& D_{\left\langle P \right\rangle} \bm{\cdot} Q \equiv  - \sum_{i=1}^n \left[ \partial_{z_i} \left(  P_i \bm{\cdot} Q  \right) \right] + \frac{d-n-1}{2} P_0   \bm{\cdot} Q . ~~~~\label{2.8}
\end{align}
Then the IBP relation \eref{eq:baikovibp1} can be written in a more compact form
\begin{align}
C \int \left\{D_{\left\langle P \right\rangle} \bm{\cdot} \frac{1}{\prod_{i=1}^n z_i^{a_i}}   \right\}  \mathcal{G}(\bm{z})^{(d-n-1)/2} \prod_{i=1}^n dz_i    . \label{eq:baikovibp2}
\end{align}
The syzygy module is a linear space with a basis of generators\footnote{For one-loop integrals, it has been proved that the number of generators is exactly
the number of propagators.}
\begin{align}
\{ \bm{e}_1 ,\cdots, \bm{e}_n  \}~~~~\label{2.10}
\end{align}
and the general solution of \eref{eq:gramsyz} can be written as $ \left\langle P \right\rangle  = \sum_{i=1}^n f_i \bm{e}_{i} $ with $f_i$s being arbitrary polynomials of $z_i$.

Another well-known phenomenon in IBP relation is the changing of power of propagators in \eref{eq:baikovibp2}.
The power  can be increased or
decreased. Among them, only
$\partial_{z_i} z_i^{-a_i}$
 increases the power. To avoid the increase, we can do a similar thing by requiring $ \left\langle P \right\rangle $ in \eref{2.8} to be the module generated by the following basis

\begin{align}
&\bm{d}_1= \{z_1,0,\cdots,0,0\} \nn\\
&\bm{d}_2= \{0,z_2,\cdots,0,0\} \nn\\
&\cdots\nn\\
&\bm{d}_n= \{0,0,\cdots,z_n,0\} \nn\\
&\bm{d}_{n+1}= \{0,0,\cdots,0,1\}.~~~~\label{2.11}
\end{align}

Up to now, we have two  modules: one is given by \eref{eq:gramsyz} and avoids the dimension shift of space-time, while another is given by \eref{2.11} and  avoids the increase of power of propagators. If we want to avoid both things, we just take the  intersection of the above two modules, i.e.,  $\{\bm{h}_i \} \equiv \{\bm{e}_i \} \cap \{\bm{d}_i \}$. Notice that the syzygy module \eref{eq:gramsyz} and the module intersection $\bm{h}_i$ can be solved by computational algebraic geometry \cite{Zhang:2016kfo}, and in this work, we use the package Singular \cite{DGPS} to do this. In the examples given
in this paper, it takes only seconds or even less to finish the computation. The syzygy of Gram determinant can also easily be obtained by Laplace expansion of the determinant \cite{Bohm:2017qme}.

\section{The method}\label{section-3}

In this paper, we will re-investigate the reduction problem for general one-loop integrals, i.e., with arbitrary tensor structure and arbitrary power of propagators. As one can see, these two different reductions, i.e., the tensor reduction (TR) and denominator reduction (DR) can be
treated uniformly by module intersection method \cite{Larsen:2015ped,Boehm:2020zig}.

As pointed out in several papers \cite{Feng:2021enk,Hu:2021nia,Feng:2022uqp,Feng:2022iuc,
Feng:2022rwj,Feng:2022rfz}, arbitrary tensor structure can be compactly organized using an auxiliary vector $R$. Thus
for TR of one-loop n-point integrals, we enlarge the set of propagators given in \eref{2.1}
by adding one new propagator, i.e.,
\bea
&&z_1=l^2-m_1^2,\ \ \  z_2=(l+p_1)^2-m_2^2,\ \ \  z_3=(l+p_1+p_2)^2-m_3^2, \cdots\ \nn\\
&&z_{n}=(l+p_1+\dots+p_{n-1})^2-m_{n}^2,\ \ \ z_{n+1}=l\cdot R.~~~\label{3.1}
\eea
with the power of $z_{n+1}$ to be non-positive integer.
Now the  Baikov representation becomes\footnote{By \eref{2.2}, the $C$ of \eref{3.2} will depend on $R$ also, but it will not influnce  IBP relations derived later.}
\begin{align}
\text{I}_{a_1,a_2,\cdots,a_{n},a_{n+1}} &=  C \int  \mathcal{G}(\bm{z})^{(d-n-2)/2}\frac{dz_i}{\prod_{i=1}^{n+1}  z_i^{a_i}} \nn\\
\mathcal{G}(\bm{z}) &= G(l,p_1,\cdots,p_{n-1},R)~~~\label{3.2}
\end{align}
Let us denote the basis of syzygy module corresponding to
\begin{align}
\left\langle \partial_{z_1} \mG ,\cdots, \partial_{z_{n+1}} \mG ,\mG \right\rangle~~~\label{3.3}
\end{align}
as $\{\bm{e}_i\}$, while the basis of another module is $\{\bm{d}_i\}$ with
\begin{align}
&\bm{d}_1= \{z_1,0,\cdots,0,0,0\} \nn\\
&\cdots\nn\\
&\bm{d}_n= \{0,0,\cdots,z_n,0,0\} \nn\\
&\bm{d}_{n+1}= \{0,0,\cdots,0,1,0\} \nn\\
&\bm{d}_{n+2}= \{0,0,\cdots,0,0,1\}~~~\label{3.4}
\end{align}
After obtained the module intersection $\{\bm{h}_i \} \equiv \{\bm{e}_i \} \cap \{\bm{d}_i \}$, we use  elements in $\{\bm{h}_i \}$ to generate differential operators as in \eref{2.8} and produce corresponding IBP relations like this\footnote{It is possible that the IBP relation can not be written to the form \eref{eq:iterTR}. For such a situation, we just throw away this IBP relation. }:
\begin{align}
&\text{I}_{\bm{a},-r_{max}} = \sum_{j=1}^m c_j \text{I}_{\bm{a},-r_{max}+j}  +  l.p.p.t. , \label{eq:iterTR}
\end{align}
with $r_{max}>0$ and $a_i>0$ for $i<n$, where the l.p.p.t. denote terms with lower power of propagators. More explicitly, we say $\text{I}_{\bm{b},-r_b}$ is a l.p.p.t. corresponding to $\text{I}_{\bm{a},-r_a}$,   if it satisfies $b_i\leq a_i$ for all $i\leq n$, and $\sum_i^n b_i < \sum_i^n a_i$.
Notice that when propagators' power is
\begin{align}
\{\bm{a},-r\} = \{1,\cdots,1,-r\},~~~\label{3.6}
\end{align}
the l.p.p.t. are all terms of sub-sectors.

To have a nice tensor reduction relation, there are some requirements for the \eref{eq:iterTR}. Firstly, the sign of power $a_{n+1}$ of $z_{n+1}$ indicates it is a
numerator or a denominator. Since we want to discuss the tensor reduction,
$a_{n+1}$ should be a non-positive integer and relation \eref{eq:iterTR} should not
include any term with $a_{n+1}$ positive. Thus for any $r_{max}>0$, we should require
\begin{align}
c_j=0\ \ \  \text{when}\ \  j>r_{max},  \label{eq:itercond}
\end{align}
Secondly, no $c_j$ becomes infinity for any $r_{max}>0$ for \eref{eq:iterTR} to be well defined.

For the reduction of propagators with arbitrary powers (i.e., the DR), the idea is similar. Without loss of generality, let us consider how to reduce the general power $a_n$ of the $n$-th propagator to one. With propagators given by
\begin{align}
&z_1=l^2-m_1^2,\ \ \  z_2=(l+p_1)^2-m_2^2,\ \ \  z_3=(l+p_1+p_2)^2-m_3^2, \cdots\ \nn\\
&z_{n}=(l+p_1+\dots+p_{n-1})^2-m_{n}^2.~~~~\label{3.8}
\end{align}
the Baikov representation is given by
\begin{align}
\text{I}_{a_1,a_2,\cdots,a_{n}} &=  C \int  \mathcal{G}(\bm{z})^{(d-n-1)/2}\frac{dz_i}{\prod_{i=1}^{n}  z_i^{a_i}} \nn\\
\mathcal{G}(\bm{z}) &= G(l,p_1,\cdots,p_{n-1}).~~~~\label{3.9}
\end{align}
Noting that for $\mathcal{G}$, when writing using momentum variables, it is the same as the one given in \eref{3.2}. However, when writing using the $z$ variables, the linear form in \eref{3.1} and the quadratic form in \eref{3.8} do make some differences.
Again first we find the syzygy module for relation \eref{eq:gramsyz} to avoid the shift of the space-time dimension. However, the second  module will be a little different from the one
given in \eref{2.11}. More explicitly,  generators become
\begin{align}
&\bm{d}_1= \{z_1,0,\cdots,0,0,0\} \nn\\
&\cdots\nn\\
&\bm{d}_{n-1}= \{0,0,\cdots,z_{n-1},0,0\} \nn\\
&\bm{d}_{n}= \{0,0,\cdots,0,1,0\} \nn\\
&\bm{d}_{n+1}= \{0,0,\cdots,0,0,1\},~~~\label{3.10}
\end{align}
where the $\bm{d}_{n}$ is different. The reason is that now
we do not ask to avoid the increase of the power of the $n$-th propagator.
Finding the module intersection $\{\bm{h}_i \} \equiv \{\bm{e}_i \} \cap \{\bm{d}_i \}$ we will get IBP relations of the form
\begin{align}
&\text{I}_{\bm{a},a_{n,max}} = \sum_{j=1}^m c_j \text{I}_{\bm{a},a_{n,max}-j}  +  l.p.p.t. ,\label{eq:iterDR}
\end{align}
where no $c_j$ becomes infinity for any $a_{n,max}>1$.
This equation is similar to \eref{eq:iterTR}, but with the following difference:  in \eref{eq:iterTR} it is the smallest power $-r_{max}$ at the left-hand side while in \eref{eq:iterDR} it is the maximum  power $a_{n,max}$ at the left-hand side.
Another difference is that in \eref{eq:iterDR} we do not need to require $c_j=0$ when $j>a_{n,max}$ since now it becomes the tensor of the sub-sector.

Before ending this section, let us emphasize that the reason we are able to treat
the TR and DR uniformly using module intersection is following two key points.
First, we have introduced the auxiliary vector $R$ to represent all tensor structures\footnote{We want to emphasize that introducing $R$  is different from introducing irreducible scalar products in usual IBP method. For later, if there are $m$ ISP's we need to introduce $m$ factors, but for $R$, we need to just introduce one for each loop momentum.}. Secondly, we are not write down all
relations coming from module intersection, but select minimum ones, which have nice property, i.e., giving iteration at the sector-level. The meaning of this point will be clear by examples in the Sec.\ref{sec:example}.

\section{Pedagogical examples}\label{sec:example}
Having discussed the method in the previous section, we will present two examples to demonstrate our method.
\subsection{tensor reduction of bubbles} \label{sec:example1}
In this subsection, we consider the tensor reduction of  one-loop bubble integrals. The propagators are
\begin{align}
 &z_1=l^2-m_1^2,\ \ \  z_2=(l+p_1)^2-m_2^2,\ \ \ z_{3}=l\cdot R.
\end{align}
and the Gram determinant in Baikov representation is
\begin{align}
\mG=\det\left(
\begin{array}{ccc}
m_1^2+z_1 & \frac{1}{2} \left(-m_1^2+m_2^2-p_1^2-z_1+z_2\right) & z_3 \\
\frac{1}{2} \left(-m_1^2+m_2^2-p_1^2-z_1+z_2\right) & p_1^2 & R\cdot p_1 \\
z_3 & R\cdot p_1 & R^2 \\
\end{array}
\right).~~~\label{4.2}
\end{align}
Using the expression in \eref{4.2} we can find
\begin{align}
&\partial_{z_1} \mG= \frac{1}{2} R^2 \left(-m_1^2+m_2^2+p_1^2-z_1+z_2\right)-z_3 R\cdot p_1-\left(R\cdot p_1\right)^2 \nn\\
&\partial_{z_2} \mG=\frac{1}{2} R^2 \left(m_1^2-m_2^2+p_1^2+z_1-z_2\right)+z_3 R\cdot p_1 \nn\\
&\partial_{z_3} \mG=-R\cdot p_1 \left(m_1^2-m_2^2+p_1^2+z_1-z_2\right)-2 p_1^2 z_3~~~\label{4.3}
\end{align}
and the solutions of equation
\begin{align}
\sum_i^3  (P_i  \partial_{z_i} \mG ) + P_0 \mG=0  ~~~\label{4.4}
\end{align}
can be solved by the syzygy module with three generators
\begin{align}
\{\bm{e}_{i}\}  = \left(
\begin{array}{cccc}
	2 z_3 & 2 \left(R\cdot p_1+z_3\right) & R^2 & 0 \\
	m_1^2+m_2^2-p_1^2+z_1+z_2 & 2 \left(m_2^2+z_2\right) & R\cdot p_1+z_3 & -2 \\
	-2 \left(m_2^2-p_1^2+z_2\right) & m_1^2-3 m_2^2-p_1^2+z_1-3 z_2 & -2 R\cdot p_1-z_3 & 2 \\
\end{array}
\right).~~~\label{4.5}
\end{align}
Meanwhile, the module $\{\bm{d}_i\}$ is generated by
\begin{align}
 \text{DM}[z_1,z_2,1,1],~~~\label{4.6}
\end{align}
where the DM denotes the diagonal matrix.
The module intersection of them is given by $\{\bm{h}_{i}\}$ with $10$ basis as polynomials of variables $\left\{z_1,z_2,z_3,m_1^2,m_2^2,p_1^2,R\cdot p_1,R^2\right\}$\footnote{The reason not using $\left\{z_1,z_2,z_3\right\}$ is explained in \cite{Bendle:2019csk}.} when using Singular \cite{DGPS}. Among them, the one with the lowest total power of $z_i$ is given by
\begin{align}
&\bm{h}_{1,1}  =
2 z_1 \left(R\cdot p_1 \left(m_1^2-m_2^2+p_1^2+z_1-z_2\right)+2 p_1^2 z_3\right) \nn\\
&\bm{h}_{1,2}  = 2 z_2 \left(R\cdot p_1 \left(m_1^2-m_2^2+p_1^2+z_1-z_2\right)+2 p_1^2 z_3\right) \nn\\
&\bm{h}_{1,3}  = R^2 \left(-p_1^2 \left(2 m_1^2+2 m_2^2+z_1+z_2\right)+\left(m_1^2-m_2^2\right) \left(m_1^2-m_2^2+z_1-z_2\right)+\left(p_1^2\right)^2\right) \nn\\
&+2 \left(\left(2 m_1^2+z_1\right) \left(R\cdot p_1\right)^2+z_3 R\cdot p_1 \left(2 m_1^2-2 m_2^2+2 p_1^2+z_1-z_2\right)+2 p_1^2 z_3^2\right) \nn\\
&\bm{h}_{1,4}  = -4 \left(R\cdot p_1 \left(m_1^2-m_2^2+p_1^2+z_1-z_2\right)+2 p_1^2 z_3\right). \label{eq:moduleintex1}
\end{align}
Using \eref{eq:moduleintex1}, one can check that the IBP relation generated by
$D_{\left\langle \bm{h}_{1} \right\rangle}$ acting on $\text{I}_{a_1,a_2,-r-1}$ can be rewritten as
\begin{align}
&\text{I}_{a_1,a_2,-r} =\frac{1}{4 p_1^2 \left(a_1+a_2-r -d+1\right)}\times  \nn\\
&\left[-2 \left(m_1^2-m_2^2+p_1^2\right) R\cdot p_1 \left(a_1+a_2-d-2 r+2  \right) \text{I}_{a_1,a_2,-(r-1)}  \right. \nn\\
& +(r-1) \left(4 m_1^2 \left(R\cdot p_1\right)^2-2 m_1^2 R^2 \left(m_2^2+p_1^2\right)+R^2 \left(m_2^2-p_1^2\right)^2+m_1^4 R^2\right)\text{I}_{a_1,a_2,-(r-2)} \nn\\
&\left. +  l.p.p.t. \right], \label{eq:exTR}
\end{align}
where
\begin{align}
 l.p.p.t.  =  &-2 R\cdot p_1 \left(a_1+a_2-r -d +1\right)  \text{I}_{a_1-1,a_2,-(r-1)}    \nn\\
 &+2 R\cdot p_1 \left(a_1+a_2-r -d+1\right) \text{I}_{a_1,a_2-1,-(r-1)}   \nn\\
 &  -(r-1) R^2  \left(m_1^2-m_2^2+p_1^2\right) \text{I}_{a_1,a_2-1,-(r-2)} \nn\\
 &  +(r-1) \left(R^2 \left(m_1^2-m_2^2-p_1^2\right)+2 \left(R\cdot p_1\right)^2\right)  \text{I}_{a_1-1,a_2,-(r-2)}.~~~\label{4.9}
\end{align}
Result \eref{eq:exTR} is the recursive relation we are looking for. Notice that when $r=1$, the coefficient of $\text{I}_{a_1,a_2,-(r-2)}=\text{I}_{a_1,a_2,1}$ is zero by  the $r-1$ factor, which satisfies the condition (\ref{eq:iterTR}). Since $r$ is the
rank of tensor in the numerator, the relation \eref{eq:exTR} tells us that the integrals of tensor rank $r$ can be written as the sum of integrals of tensor rank $(r-1)$ and $(r-2)$ with proper rational coefficients. This kind of relations has been firstly  observed in \cite{Feng:2022iuc} for the case $a_1=a_2=1$ and then  has been proved in \cite{Feng:2022rfz}. Here we give a simple derivation using module intersection with generalization to arbitrary power of propagators, as well as given the analytic l.p.p.t part missed in \cite{Feng:2022iuc,Feng:2022rfz}.  It is obvious that applying this kind of second-order iterative relation, one  can immediately reduce any tensor integrals of this sector to scalar integrals of the same sector and tensor integrals of sub-sectors.

Another interesting property of relation (\ref{eq:exTR}) is that in \eref{eq:exTR} the $a_1,a_2$ of bubbles are invariant, while for l.p.p.t. in \eref{4.9}, $(a_1+a_2)$ has changed only by minus one.
This observation  will be explained in Sec.\ref{sec:fomula}.

Before ending this subsection, let us emphasize that to deal with tensor reduction of bubble topology, we have required $a_3=-r<0$. If we consider the case $a_3>0$, we will get the triangle topology, and the relation becomes IBP relation for triangles, but with the third propagator is not the standard quadratic one. In next subsection, we will consider triangle topology with the standard Feynman propagators, thus it will be useful to compare results in these two subsections.

\subsection{DR reduction of triangle}\label{sec:example2}

In this subsection, we discuss the reduction of triangles with arbitrary powers for propagators. The propagators are
\begin{align}
&z_1=l^2-m_1^2,\ \ \  z_2=(l+p_1)^2-m_2^2,\ \ \ z_{3}=(l+p_1+p_2)^2-m_3^2.
\end{align}
and the corresponding Gram determinant $\mG$ in Baikov representation is
\begin{align}
\det\left(
\begin{array}{ccc}
m_1^2+z_1 & \cdots & \cdots \\
\frac{1}{2} \left(-m_1^2+m_2^2-p_1^2-z_1+z_2\right) & p_1^2 & \cdots \\
\frac{1}{2} \left(-m_2^2+m_3^2-p_2^2-2 p_1 \cdot p_2-z_2+z_3\right) & p_1 \cdot p_2 & p_2^2 \\
\end{array}
\right)
\end{align}
where the $\cdots$ denote the terms, which can be obtained by symmetry. With
\begin{align}
&\partial_{z_1} \mG= \frac{1}{2} \left(p_2^2 \left(-m_1^2+m_2^2+p_1^2-z_1+z_2\right)+p_1 \cdot p_2 \left(m_2^2-m_3^2+p_2^2+z_2-z_3\right)\right), \nn\\
&\partial_{z_2} \mG=\frac{1}{2} \left(p_1 \cdot p_2 \left(m_1^2-2 m_2^2+m_3^2-p_1^2-p_2^2+z_1-2 z_2+z_3\right)+m_3^2 p_1^2+m_1^2 p_2^2 \right. \nn\\
& \left. -m_2^2 \left(p_1^2+p_2^2\right)+p_2^2 z_1-p_1^2 z_2-p_2^2 z_2+p_1^2 z_3-2 \left(p_1 \cdot p_2\right)^2\right), \nn\\
&\partial_{z_3} \mG=\frac{1}{2} \left(p_1^2 \left(m_2^2-m_3^2+p_2^2+z_2-z_3\right)+p_1 \cdot p_2 \left(-m_1^2+m_2^2+p_1^2-z_1+z_2\right)\right).
\end{align}
one can solve
\begin{align}
\sum_i^3  (P_i  \partial_{z_i} \mG ) +  P_0 \mG=0  ,\label{eq:syzp3}
\end{align}
with the basis  $\{\bm{e}_{i}\}$ of the syzygy module is
\begin{align}
\left(
\begin{array}{cccc}
m_1^2+m_3^2-s+z_1+z_3 & m_2^2+m_3^2-p_2^2+z_2+z_3 & 2 \left(m_3^2+z_3\right) & -2 \\
m_1^2+m_2^2-p_1^2+z_1+z_2 & 2 \left(m_2^2+z_2\right) & m_2^2+m_3^2-p_2^2+z_2+z_3 & -2 \\
2 \left(m_1^2+z_1\right) & m_1^2+m_2^2-p_1^2+z_1+z_2 & m_1^2+m_3^2-s+z_1+z_3 & -2 \\
\end{array}
\right)
\end{align}
where $s=(p_1+p_2)^2$. 
Another module is generated by (see \eref{3.10})
\begin{align}
\{\bm{d}_{i}\}  = \text{DM}[z_1,z_2,1,1],
\end{align}
From them, we can compute $\{\bm{h}_i\}$ as the module intersection of them. Among them,  the one  with the lowest total power of $z_i$ is given by
\begin{align}
&\bm{h}_{1,1}  = 2 z_1 \left(p_1^2 \left(m_2^2-m_3^2+p_2^2+z_2-z_3\right)+p_1 \cdot p_2 \left(-m_1^2+m_2^2+p_1^2-z_1+z_2\right)\right), \nn\\
&\bm{h}_{1,2}  = 2 z_2 \left(p_1^2 \left(m_2^2-m_3^2+p_2^2+z_2-z_3\right)+p_1 \cdot p_2 \left(-m_1^2+m_2^2+p_1^2-z_1+z_2\right)\right), \nn\\
&\bm{h}_{1,3}  = -2 \left(m_1^2 p_2^2 z_1-m_1^2 p_2^2 z_2-m_3^2 p_1^2 z_2+2 \left(p_1 \cdot p_2\right)^2 \left(2 m_2^2+z_2\right)+2 m_3^2 p_1^2 z_3 \right. \nn\\
& +m_2^2 \left(-2 m_3^2 p_1^2-2 m_1^2 p_2^2+p_1^2 z_2-2 p_1^2 z_3-p_2^2 z_1+p_2^2 z_2\right)  \nn\\
&+p_1 \cdot p_2 \left(-m_2^2 \left(2 m_3^2-2 p_1^2-2 p_2^2+z_1-2 z_2+2 z_3\right)-m_1^2 \left(2 m_2^2-2 m_3^2+2 p_2^2+z_2-2 z_3\right) \right. \nn\\
& \left. -2 m_3^2 p_1^2+m_3^2 z_1-m_3^2 z_2+2 m_2^4-p_2^2 z_1+p_1^2 z_2+p_2^2 z_2-2 p_1^2 z_3+2 p_1^2 p_2^2+z_1 z_3-z_2 z_3\right)  \nn\\
& +m_1^4 p_2^2-2 m_1^2 p_1^2 p_2^2+m_3^4 p_1^2-2 m_3^2 p_1^2 p_2^2+m_2^4 \left(p_1^2+p_2^2\right)+p_1^2 z_3^2-p_1^2 p_2^2 z_1-2 p_1^2 p_2^2 z_3\nn\\
&\left. -p_1^2 z_2 z_3+p_1^2 p_2^4+p_1^4 p_2^2\right), \nn\\
&\bm{h}_{1,4}  = 4 p_1 \cdot p_2 \left(m_1^2-m_2^2-p_1^2+z_1-z_2\right)-4 p_1^2 \left(m_2^2-m_3^2+p_2^2+z_2-z_3\right).   \label{eq:moduleintex2}
\end{align}
Using \eref{eq:moduleintex2}
the action of $D_{\left\langle \bm{h}_{1} \right\rangle}$  on $\text{I}_{a_1,a_2,a_3-1}$ gives the wanted IBP relation
\begin{align}
&\text{I}_{a_1,a_2,a_3} =  \frac{1}{Q_1}\times \left[ -p_1^2 \left(a_1+a_2+a_3 -d -1\right) \text{I}_{a_1,a_2,a_3-2} \right.\nn\\
&  \left(2 a_3 +a_1+a_2 -d -2\right) \left(p_1^2 \left(m_2^2-m_3^2+p_2^2\right)+p_1 \cdot p_2 \left(-m_1^2+m_2^2+p_1^2\right)\right) \text{I}_{a_1,a_2,a_3-1}   \nn\\
&\left.+ l.p.p.t.\right] \label{eq:exDR}
\end{align}
where
\begin{align}
l.p.p.t.  &=  \left(a_3-1\right) \left(p_2^2 \left(-m_1^2+m_2^2+p_1^2\right)+p_1 \cdot p_2 \left(m_2^2-m_3^2+p_2^2\right)\right) \text{I}_{a_1-1,a_2,a_3}    \nn\\
&-\left(a_3-1\right) \left(p_1 \cdot p_2 \left(-m_1^2+2 m_2^2-m_3^2+p_1^2+p_2^2\right)-m_3^2 p_1^2-m_1^2 p_2^2+m_2^2 \left(p_1^2+p_2^2\right) \right. \nn\\
&\left. +2 \left(p_1 \cdot p_2\right)^2\right) \text{I}_{a_1,a_2-1,a_3}   \nn\\
&  -p_1 \cdot p_2 \left(a_1+a_2+a_3 -d -1\right) \text{I}_{a_1-1,a_2,a_3-1} \nn\\
&   +\left(p_1^2+p_1 \cdot p_2\right)\left(a_1+a_2+a_3 -d -1\right) \text{I}_{a_1,a_2-1,a_3-1} \nn\\
Q_1&=\left(a_3-1\right) \left(m_1^4 p_2^2-2 m_1^2 p_1^2 p_2^2+4 m_2^2 \left(p_1 \cdot p_2\right)^2+m_3^4 p_1^2-2 m_3^2 p_1^2 p_2^2 +m_2^4 \left(p_1^2+p_2^2\right) \right.\nn\\
&\left.-2 p_1 \cdot p_2 \left(m_1^2-m_2^2-p_1^2\right) \left(m_2^2-m_3^2+p_2^2\right)-2 m_2^2 \left(m_3^2 p_1^2+m_1^2 p_2^2\right)+p_1^2 p_2^4+p_1^4 p_2^2 \right).
~~~\label{4.18}
\end{align}
Notice that when $a_3=2$, the coefficient of $\text{I}_{a_1,a_2,a_3-2}=\text{I}_{a_1,a_2,0}$ is a term of sub-sector (here is the bubble topology). Also for $a_3=1$ we don't need to apply this relation  for DR since it is already the final goal we want to achieve. The relation \eref{eq:exDR} is also a second-order iterative relation relating $\text{I}_{a_1,a_2,a_3}$ to $\text{I}_{a_1,a_2,a_3-1}$, $\text{I}_{a_1,a_2,a_3-2}$ and l.p.p.t.. Similar to \eref{eq:exTR},
the $a_1, a_2$ of triangles in \eref{eq:exDR}
are invariant, while for l.p.p.t. in \eref{4.18}, $(a_1 + a_2)$ has changed only by minus one.

Thus applying the relation  \eref{eq:exDR} iteratively one can immediately reduce any higher power of $z_3$ to one. Similar iterative relations for reducing the power of other propagators can be obtained.  Combining them, we can reduce integrals with arbitrary high power of propagators in this sector.

\section{Uniform formula for general one-loop reduction}\label{sec:fomula}

The method laid out in section \ref{section-3} and related computations done in section
\ref{sec:example} look fresh, but maybe not so surprising. However, as we will show in this section, for one-loop integrals, we can write down explicit recursive relations for both TR and DR uniformly for any general $n$-point one-loop integrals. In other words, we have solved IBP relations analytically.

The key of our method is to select particular elements in  module intersection $\{\bm{h}_i\}$.
Let's make an observation for two examples in section \ref{sec:example}, i.e.,  both results (\ref{eq:moduleintex1}) and (\ref{eq:moduleintex2}), we find that
\begin{align}
\{\bm{h}_{1,1},\bm{h}_{1,2},\bm{h}_{1,4}\}=C  \times \{z_1 \partial_{z_3} \mG,z_2 \partial_{z_3} \mG,-2 \partial_{z_3} \mG\}
\end{align}
and then by (\ref{eq:syzp3}), we have
\begin{align}
 \bm{h}_{1,3} = C \times \left( 2\mG  - z_1 \partial_{z_1} \mG - z_2 \partial_{z_2} \mG \right) .
\end{align}
This pattern indicates that for DR of the  $N$-th propagator of the $N$-point one-loop integrals or the TR  of $(N-1)$-point one-loop integrals (where the $(R\cdot \ell)$ has been considered as the $N$-th propagator), the wanted element in the  intersection module is given by
\begin{align}
&P_{ui}=z_i \partial_{z_N} \mG  ~~~\text{for }~ 1\leq i \leq N-1, \nn\\
&P_{uN}=2\mG-\sum_{i=1}^{N-1} z_i \partial_{z_i} \mG,~~~~~~P_{u0}= -2 \partial_{z_N} \mG  .~~~\label{5.3}
\end{align}
It  is easy to check that \eref{5.3} satisfies  \eref{eq:gramsyz} and belongs to the module
\begin{align}
\{\bm{d}_{i}\}  = \text{DM}[z_1,z_2,\cdots,z_{N-1},1,1],
\end{align}
then the  differential operator $D_{\langle P_{u} \rangle}$ gives the generic iterative relation for both TR and DR of general one-loop integrals.


To write down explicit relation, let us compute $P_u$ in \eref{5.3}.
For  one-loop integrals, Gram determinant $\mG$ is always quadratic polynomial of $z_i$s
\begin{align}
&\mG(\bm{z})=   \sum_{i,j\leq i} C_2^{(ij)} z_i z_j  + \sum_{i} C_1^{(i)} z_i + C_0~~~\label{5.5}
\end{align}
where
\begin{align}
&C_2^{(ii)} = \partial_{z_i}^2\mG /2,~~~ C_1^{(i)} = \left(\partial_{z_i}\mG \right)|_{\bm{z}=0},~~~ C_0 = \mG|_{\bm{z}=0},\nn\\
&C_2^{(ij)} = C_2^{(ji)} = \partial_{z_i}\partial_{z_j}\mG  ~~~~~\text{for}~~ j\neq i  .~~~\label{5.6}
\end{align}
This leads to
\begin{align}
&\partial_{z_i}  \mG = 2  C_2^{(ii)} z_i  + \sum_{j\neq i} C_2^{(ij)} z_j +   C_1^{(i)} , \nn\\
& \sum_{i=1}^{N} z_i \partial_{z_i}  \mG = 2 \sum_{i,j\leq i} C_2^{(ij)} z_i z_j +  \sum_{i} C_1^{(i)} z_i, \nn\\
&P_{uN}=  z_N \partial_{z_N}\mG+     \sum_{i} C_1^{(i)} z_i + 2 C_0  .~~~\label{5.7}
\end{align}
The rewriting of $P_{uN}$ in \eref{5.7} tells us that $P_{uN}$ depends on $z_i, i=1,...,N-1$ only linearly.

Combining $P_{ui}$ in \eref{5.3} and $\partial_{z_i}  \mG$ in \eref{5.7}, one can see that
the power of $z_i$s  will  lead  all $\text{I}_{a^\prime_1,\cdots,a^\prime_{N-1},a^\prime_{N}}$
to satisfy $0\leq \sum_{i=1}^{N-1}a_i - \sum_{i=1}^{N-1}a^\prime_i \leq 1$. To show that, let us do the following explicit computations.


For $i\leq N-1$, carrying out
\begin{align}
&-D_{P_{ui}}\bm{\cdot} \frac{1}{\prod_{i=1}^{N} z_i^{a_i}} = -\partial_{z_i} \left( \frac{z_i \partial_{z_N} \mG}{\prod_{i=1}^{N} z_i^{a_i}}  \right),~~~\label{5.8}
\end{align}
we find
\begin{align}
&(a_i-1) \left(2 C_{2}^{(N N)} \text{I}_{\cdots,a_N-1} +  C_{1}^{(N)} \text{I}_{\cdots,a_N} \right) \nn\\
+& (a_i-2) C_{2}^{(N i)} \text{I}_{\cdots,a_i-1,\cdots,a_N} + (a_i-1)   \sum_{j=1,j\neq i}^{N-1} C_{2}^{(N j)}  \text{I}_{\cdots,a_j-1,\cdots,a_N} .~~~\label{5.9}
\end{align}
For $i=N$, action of $D_{P_N}$  gives
\begin{align}
&2\left[ (a_N-2) C_{2}^{(N N)} \text{I}_{\cdots,a_N-1} + (a_{N}-1)  C_{1}^{(N)}  \text{I}_{\cdots,a_N} + a_N  C_{0}  \text{I}_{\cdots,a_N+1} \right] \nn\\
 &+\left[  (a_N-1)   \sum_{j\neq N} C_{2}^{(Nj)} \text{I}_{\cdots,a_j-1,\cdots,a_N}  + a_N   \sum_{j\neq N} C_{1}^{(j)} \text{I}_{\cdots,a_j-1,\cdots,a_N-1} \right] .~~~\label{5.10}
\end{align}
Finally action of $D_{P_0}$  gives
\begin{align}
& -  \left( d-N-1 \right)  \left(  C_1^{(N)} \text{I}_{\cdots,a_N} + 2 C_2^{(NN)} \text{I}_{\cdots,a_N-1}+\sum_{j=1}^{N-1} C_2^{(Nj)} \text{I}_{\cdots,a_j-1,\cdots,a_N}  \right) .~~~\label{5.11}
\end{align}
Combining all together, we have
\begin{align}
&2 a_N C_{0}  \text{I}_{\cdots,a_N+1} +\left(  \sum_{i=1}^{N-1} a_i +2  a_N - d  \right)  C_1^{(N)} \text{I}_{\cdots,a_N} \nn\\
+&\left(  \sum_{i=1}^{N} a_i - d  \right)  2  C_2^{(NN)} \text{I}_{\cdots,a_N-1}   +  l.p.p.t.=0~~~\label{5.12}
\end{align}
where
\begin{align}
l.p.p.t. = &\left(\sum_{i=1}^{N} a_i  -d \right) \sum_{j=1}^{N-1} C_2^{(Nj)} \text{I}_{\cdots,a_j-1,\cdots,a_N} +a_N \sum_{j=1}^{N-1} C_1^{(j)} \text{I}_{\cdots,a_j-1,\cdots,a_N-1}~.~~\label{5.13}
\end{align}
Expressions \eref{5.12} and \eref{5.13} are our main results for this paper.

When $a_{N}<0$, it corresponds to the numerator and we should use \eref{5.12} to express $\text{I}_{\cdots,a_N-1}$ by others for the TR, while when $a_{N}>0$ it corresponds to the denominator with a higher power and we should use \eref{5.12} to express $\text{I}_{\cdots,a_N+1}$ by others for the DR. When $a_N=0$, the \eref{5.12} will give the relation between $\text{I}_{\cdots,-1}$ and $\text{I}_{\cdots,0}$, which is just the reduction of tensor with rank one.

\section{Example of degenerated case}\label{section-6}

In section \ref{sec:fomula}, we have assumed that the kinematics and masses are general.
But when kinematics and masses take some special values, the Gram determinant may be zero and we meet the degenerated situations. For these cases,
directly applying \eref{5.12} and \eref{5.13} still work, but why it works should be carefully
explained. We will show that there are three different possibilities. The first one is that
while it works, it may not give the simplest iterative relation. The second one is that while
it does not work for some situations, there is at least one situation it works. The third one is that it does not work at all, but for this case, there are no master integrals, so it does not matter. No matter which situation one meets,  directly applying the method of module intersection presented in the Sec.\ref{section-3} gives an alternative way to do reduction than using \eref{5.12} and \eref{5.13}.



Now we give an example to elaborate on the above claim.
Let us consider the  reduction of triangles
\begin{align}
&z_1=l^2,\ \ \  z_2=(l+p_1)^2,\ \ \  z_3=(l+p_1+p_2)^2,\nn\\
&z_{4}=l\cdot R, \ \ \ p_1^2=p_2^2=0.~~~~\label{6.1}
\end{align}
with specific kinematics.
As shown later, by IBP relations one can see  that all integrals in this sector can be reduced to sub-sectors.

For TR of \eref{6.1}, $\mG=G(l,p_1,p_2,R)$.  Directly applying the formula \eref{5.12} and \eref{5.13}, we will get\footnote{$R^2$ does not appear in \eref{eq:svf}
because $G(l,p_1,p_2)|_{\bm{z}=0}=0$ when using  \eref{5.12} and \eref{5.13}.}
\begin{align}
&\text{I}_{a_1,a_2,a_3,-r-1} = \frac{R \cdot p_1 \left(r \ R \cdot p_1\  \text{I}_{a_1,a_2,a_3,1-r}-\left(\sum_{i}^3 a_i-d-2 r\right) \text{I}_{a_1,a_2,a_3,-r}\right)}{\sum_{i}^3 a_i-d-r}  + l.p.p.t.\  , \label{eq:svf}
\end{align}
which is a second-order iterative relation. Using \eref{eq:svf}, one can still reduce all $\text{I}_{1,1,1,-r}$  to $\text{I}_{1,1,1,0}$. However,   $\text{I}_{1,1,1,0}$ is not a master integral in this example. To reduce
$\text{I}_{1,1,1,0}$ we should use
\eref{5.12} and \eref{5.13} for DR, which will be explained shortly.

If we calculate the module intersection follow the method in Sec.\ref{section-3}, one can find another element $\bm{h}_i$ different from the one given in \eref{5.3}, which
will give us the following iterative reduction relation
\begin{align}
\text{I}_{a_1,a_2,a_3,-r}=\frac{r R \cdot p_1 \text{I}_{a_1,a_2,a_3,1-r}}{2 a_2+2 a_3-d-r} + l.p.p.t.\ .~~~~\label{6.3}
\end{align}
Obviously, as a first-order iterative relation,  \eref{6.3} is simpler than (\ref{eq:svf}), which supports the claim of the first possibility.

For DR\footnote{Now there is no $z_4$ in \eref{6.1}.}, the Gram determinant in Baikov representation is
\begin{align}
\mG=G(l,p_1,p_2)=\frac{1}{2} p_1 \cdot p_2 \left(\left(z_1-z_2\right) \left(z_2-z_3\right)-2 p_1 \cdot p_2 z_2\right).~~~\label{6.4}
\end{align}
To determine the number of master integrals, we can consider the maximum cut of this sector in Baikov representation, i.e., setting $z_i=0$ in $\mG$. If $\mG|_{\bm{z}=0}$ is a nonzero
constant, the number of master integrals is just one by counting critical points \cite{Lee:2013hzt,Frellesvig:2019kgj,Chen:2022lzr}. But if $\mG|_{\bm{z}=0}=0$,
there is no master integral in this sector. It is easy to see that
$\mG|_{\bm{z}=0}=0$ in \eref{6.4}, thus $\text{I}_{1,1,1}$ can be reduced to sub-sectors.

One can check that if we regard $z_1$ or $z_3$ as the $z_N$ and apply \eref{5.12}, the $C_0, C_1^{(N)}, C_{2}^{(NN)}$ are all zero and only l.p.p.t. is left. In other words, for these cases, \eref{5.12} does not produce the wanted relation for reduction purpose.
However, if we regard $z_2$ as the $z_N$, the iterative relation reduce to first-order due to $C_0=0$, $ C_1^{(2)}\neq 0$ and $ C_2^{(22)}\neq 0$. Using it, we can reduce
$I_{a_1,a_2,a_3,0}$ (including the $I_{1,1,1,0}$ discussed in previous paragraph) to sub-sectors $I_{a_1,0,a_3,0}$ and others.
The $a_1$ and $a_3$ are left to be reduced by DR of \eref{5.12} in the sub-sector. So, as pointed out for the second possibility,  although using \eref{5.12} for the DR does not work for $z_1$ and $z_3$, there is $z_2$ it works.

Nevertheless, even if we regard $z_3$ as the $z_N$, we can do similar module computation proposed in Sec.\ref{section-3}. The syzygy module is generated by
\begin{align}
&\{\bm{e_i}\}= \left(
\begin{array}{cccc}
z_2 & z_2 & z_3 & -1 \\
2 p_1 \cdot p_2-z_3 & -z_3 & z_2-2 z_3 & 1 \\
-2 p_1 \cdot p_2+z_1-z_2+z_3 & z_3 & z_3 & -1 \\
4 p_1 \cdot p_2+z_2-2 z_3 & z_1-2 z_3 & -2 p_1 \cdot p_2+z_1-2 z_3 & 1 \\
\end{array}
\right) , \label{6.5}
\end{align}
while the  element in the intersection module is taken to be
\bea \aP=\langle z_1,z_2,z_2,1 \rangle. \label{6.6}\eea
Using \eref{6.6}, the iterative relation is
\begin{align}
&\text{I}_{a_1,a_2,a_3,0}= -\frac{2 a_3  \text{I}_{a_1,a_2-1,a_3+1,0}}{2 a_1+2 a_2-d}~~~~\label{6.7}
\end{align}
where the $(a_1+a_2)$ has been reduced by one at the right-hand side, which is just a l.p.p.t..
Although this relation raises the power of $z_3$ in denominator,  it will lower $a_2$ to $0$ finally (i.e., reduce to sub-sector), so it shows that
method of module intersection still works.

From this example, it is easy to see that only for more degenerated case, i.e., all $C_2^{(ii)}$, $C_1^{(i)}$ and $C_0$ in $\mG$  equal zero,  \eref{5.12} can not reduce integrals in this sector to sub-sectors. But for this situation, all sub-sectors have no master integrals by counting the critical points. To be more explicitly, the number of master integrals in the corresponding sector is equal to the number of solutions to the equations
\begin{align}
0=\partial_{z_i}\log\left(\mG^{(d-n-1)/2}\right)|_{\bm{z^\prime}=0},\ \ \ \text{for all}~z_i\not \in\{\bm{z}^\prime\},  \label{misofsub}
\end{align}
Now the $\mG$ takes the form
\begin{align}
\mG=\sum_{i,j\neq i}c_{ij} z_i z_j   \label{misofsub1}
\end{align}
and the equations \eref{misofsub}  take the form
\begin{align}
\frac{C}{z_i-C_i} =0 ,  \label{misofsub2}
\end{align}
which obviously has no solution. It means that integrals in this topology are all scaleless integrals, and this topology has no master integral. People can test this conclusion by taking $p_1\cdot p_2$ to zero in \eref{6.4} and immediately find this topology to be scaleless. This is the third possibility we have mentioned.


\section{Summary and Outlook}\label{sec:conc}

In this paper, a natural extension of syzygy and module intersection has been shown that it can  uniformly  reduce one-loop integrals with arbitrary tensor (by using auxiliary vector) and propagators with high powers. With a nice observation, powerful iterative relation can
be written down even without doing explicit computations using computational algebraic geometry and module intersection. Furthermore,  in this formula, the polynomials that look messy in the traditional reduction methods are arranged themselves to Gram determinants. Such a property will not only  speed up the analytic reduction of Feynman integrals,  but also  help to investigate mathematical structures of Feynman integrals and amplitudes in the future.



It is obvious that the motivation of our study is the reduction for high loops. If the method is good, it must work well at the one-loop level. Results in this paper have demonstrate this point.
To generalize this method to high loops, although steps of computational algebraic geometry can be easily applied, some nontrivial problems will arise. There are irreducible scalar products of loop momenta and more than one master integrals for a given sector in most multi-loop integrals. This suggests that for multi-loop integrals,  elements of module intersection needed may also be more than one. More importantly, could we obtain the general reduction formulas like \eref{5.12} for multi-loops? Are there also some hidden informations or structures in the messy polynomials in the IBP relation? These problems are definitely interesting  for future exploration.


\section*{Acknowledgments}
This work is supported by  Chinese NSF funding under Grant No.11935013, No.11947301, No.12047502 (Peng Huanwu Center).

\bibliographystyle{JHEP}
\bibliography{references}

\providecommand{\href}[2]{#2}\begingroup\raggedright\begin{thebibliography}{10}

\bibitem{Chetyrkin:1981qh}
K.G.~Chetyrkin and F.V.~Tkachov, \emph{{Integration by Parts: The Algorithm to
  Calculate beta Functions in 4 Loops}},
  \href{https://doi.org/10.1016/0550-3213(81)90199-1}{\emph{Nucl. Phys. B}
  {\bfseries 192} (1981) 159}.

\bibitem{Kotikov:1990kg}
A.V.~Kotikov, \emph{{Differential equations method: New technique for massive
  Feynman diagrams calculation}},
  \href{https://doi.org/10.1016/0370-2693(91)90413-K}{\emph{Phys. Lett. B}
  {\bfseries 254} (1991) 158}.

\bibitem{Gehrmann:1999as}
T.~Gehrmann and E.~Remiddi, \emph{{Differential equations for two loop four
  point functions}},
  \href{https://doi.org/10.1016/S0550-3213(00)00223-6}{\emph{Nucl. Phys. B}
  {\bfseries 580} (2000) 485}
  [\href{https://arxiv.org/abs/hep-ph/9912329}{{\ttfamily hep-ph/9912329}}].

\bibitem{Argeri:2007up}
M.~Argeri and P.~Mastrolia, \emph{{Feynman Diagrams and Differential
  Equations}}, \href{https://doi.org/10.1142/S0217751X07037147}{\emph{Int. J.
  Mod. Phys. A} {\bfseries 22} (2007) 4375}
  [\href{https://arxiv.org/abs/0707.4037}{{\ttfamily 0707.4037}}].

\bibitem{Henn:2014qga}
J.M.~Henn, \emph{{Lectures on differential equations for Feynman integrals}},
  \href{https://doi.org/10.1088/1751-8113/48/15/153001}{\emph{J. Phys. A}
  {\bfseries 48} (2015) 153001}
  [\href{https://arxiv.org/abs/1412.2296}{{\ttfamily 1412.2296}}].

\bibitem{Henn:2013pwa}
J.M.~Henn, \emph{{Multiloop integrals in dimensional regularization made
  simple}}, \href{https://doi.org/10.1103/PhysRevLett.110.251601}{\emph{Phys.
  Rev. Lett.} {\bfseries 110} (2013) 251601}
  [\href{https://arxiv.org/abs/1304.1806}{{\ttfamily 1304.1806}}].

\bibitem{Moriello:2019yhu}
F.~Moriello, \emph{{Generalised power series expansions for the elliptic planar
  families of Higgs + jet production at two loops}},
  \href{https://doi.org/10.1007/JHEP01(2020)150}{\emph{JHEP} {\bfseries 01}
  (2020) 150} [\href{https://arxiv.org/abs/1907.13234}{{\ttfamily
  1907.13234}}].

\bibitem{Bonciani:2019jyb}
R.~Bonciani, V.~Del~Duca, H.~Frellesvig, J.M.~Henn, M.~Hidding, L.~Maestri
  et~al., \emph{{Evaluating a family of two-loop non-planar master integrals
  for Higgs + jet production with full heavy-quark mass dependence}},
  \href{https://doi.org/10.1007/JHEP01(2020)132}{\emph{JHEP} {\bfseries 01}
  (2020) 132} [\href{https://arxiv.org/abs/1907.13156}{{\ttfamily
  1907.13156}}].

\bibitem{Frellesvig:2019byn}
H.~Frellesvig, M.~Hidding, L.~Maestri, F.~Moriello and G.~Salvatori, \emph{{The
  complete set of two-loop master integrals for Higgs + jet production in
  QCD}}, \href{https://doi.org/10.1007/JHEP06(2020)093}{\emph{JHEP} {\bfseries
  06} (2020) 093} [\href{https://arxiv.org/abs/1911.06308}{{\ttfamily
  1911.06308}}].

\bibitem{Hidding:2020ytt}
M.~Hidding, \emph{{DiffExp, a Mathematica package for computing Feynman
  integrals in terms of one-dimensional series expansions}},
  \href{https://doi.org/10.1016/j.cpc.2021.108125}{\emph{Comput. Phys. Commun.}
  {\bfseries 269} (2021) 108125}
  [\href{https://arxiv.org/abs/2006.05510}{{\ttfamily 2006.05510}}].

\bibitem{Liu:2017jxz}
X.~Liu, Y.-Q.~Ma and C.-Y.~Wang, \emph{{A Systematic and Efficient Method to
  Compute Multi-loop Master Integrals}},
  \href{https://doi.org/10.1016/j.physletb.2018.02.026}{\emph{Phys. Lett. B}
  {\bfseries 779} (2018) 353}
  [\href{https://arxiv.org/abs/1711.09572}{{\ttfamily 1711.09572}}].

\bibitem{Liu:2022tji}
Z.-F.~Liu and Y.-Q.~Ma, \emph{{Automatic computation of Feynman integrals
  containing linear propagators via auxiliary mass flow}},
  \href{https://doi.org/10.1103/PhysRevD.105.074003}{\emph{Phys. Rev. D}
  {\bfseries 105} (2022) 074003}
  [\href{https://arxiv.org/abs/2201.11636}{{\ttfamily 2201.11636}}].

\bibitem{Liu:2022mfb}
Z.-F.~Liu and Y.-Q.~Ma, \emph{{Feynman integrals are completely determined by
  linear algebra}},  \href{https://arxiv.org/abs/2201.11637}{{\ttfamily
  2201.11637}}.

\bibitem{Liu:2022chg}
X.~Liu and Y.-Q.~Ma, \emph{{AMFlow: a Mathematica package for Feynman integrals
  computation via Auxiliary Mass Flow}},
  \href{https://arxiv.org/abs/2201.11669}{{\ttfamily 2201.11669}}.

\bibitem{Liu:2021wks}
X.~Liu and Y.-Q.~Ma, \emph{{Multiloop corrections for collider processes using
  auxiliary mass flow}},
  \href{https://doi.org/10.1103/PhysRevD.105.L051503}{\emph{Phys. Rev. D}
  {\bfseries 105} (2022) L051503}
  [\href{https://arxiv.org/abs/2107.01864}{{\ttfamily 2107.01864}}].

\bibitem{Liu:2020kpc}
X.~Liu, Y.-Q.~Ma, W.~Tao and P.~Zhang, \emph{{Calculation of Feynman loop
  integration and phase-space integration via auxiliary mass flow}},
  \href{https://doi.org/10.1088/1674-1137/abc538}{\emph{Chin. Phys. C}
  {\bfseries 45} (2021) 013115}
  [\href{https://arxiv.org/abs/2009.07987}{{\ttfamily 2009.07987}}].

\bibitem{Armadillo:2022ugh}
T.~Armadillo, R.~Bonciani, S.~Devoto, N.~Rana and A.~Vicini, \emph{{Evaluation
  of Feynman integrals with arbitrary complex masses via series expansions}},
  \href{https://arxiv.org/abs/2205.03345}{{\ttfamily 2205.03345}}.

\bibitem{Feng:2022uqp}
B.~Feng, T.~Li, H.~Wang and Y.~Zhang, \emph{{Reduction of general one-loop
  integrals using auxiliary vector}},
  \href{https://doi.org/10.1007/JHEP05(2022)065}{\emph{JHEP} {\bfseries 05}
  (2022) 065} [\href{https://arxiv.org/abs/2203.14449}{{\ttfamily
  2203.14449}}].

\bibitem{Laporta:2000dsw}
S.~Laporta, \emph{{High precision calculation of multiloop Feynman integrals by
  difference equations}},
  \href{https://doi.org/10.1142/S0217751X00002159}{\emph{Int. J. Mod. Phys. A}
  {\bfseries 15} (2000) 5087}
  [\href{https://arxiv.org/abs/hep-ph/0102033}{{\ttfamily hep-ph/0102033}}].

\bibitem{Smirnov:2019qkx}
A.V.~Smirnov and F.S.~Chuharev, \emph{{FIRE6: Feynman Integral REduction with
  Modular Arithmetic}},
  \href{https://doi.org/10.1016/j.cpc.2019.106877}{\emph{Comput. Phys. Commun.}
  {\bfseries 247} (2020) 106877}
  [\href{https://arxiv.org/abs/1901.07808}{{\ttfamily 1901.07808}}].

\bibitem{vonManteuffel:2012np}
A.~von Manteuffel and C.~Studerus, \emph{{Reduze 2 - Distributed Feynman
  Integral Reduction}},  \href{https://arxiv.org/abs/1201.4330}{{\ttfamily
  1201.4330}}.

\bibitem{Lee:2012cn}
R.N.~Lee, \emph{{Presenting LiteRed: a tool for the Loop InTEgrals REDuction}},
   \href{https://arxiv.org/abs/1212.2685}{{\ttfamily 1212.2685}}.

\bibitem{Klappert:2020nbg}
J.~Klappert, F.~Lange, P.~Maierh\"ofer and J.~Usovitsch, \emph{{Integral
  reduction with Kira 2.0 and finite field methods}},
  \href{https://doi.org/10.1016/j.cpc.2021.108024}{\emph{Comput. Phys. Commun.}
  {\bfseries 266} (2021) 108024}
  [\href{https://arxiv.org/abs/2008.06494}{{\ttfamily 2008.06494}}].

\bibitem{Bern:1994zx}
Z.~Bern, L.J.~Dixon, D.C.~Dunbar and D.A.~Kosower, \emph{{One loop n point
  gauge theory amplitudes, unitarity and collinear limits}},
  \href{https://doi.org/10.1016/0550-3213(94)90179-1}{\emph{Nucl. Phys. B}
  {\bfseries 425} (1994) 217}
  [\href{https://arxiv.org/abs/hep-ph/9403226}{{\ttfamily hep-ph/9403226}}].

\bibitem{Bern:1994cg}
Z.~Bern, L.J.~Dixon, D.C.~Dunbar and D.A.~Kosower, \emph{{Fusing gauge theory
  tree amplitudes into loop amplitudes}},
  \href{https://doi.org/10.1016/0550-3213(94)00488-Z}{\emph{Nucl. Phys. B}
  {\bfseries 435} (1995) 59}
  [\href{https://arxiv.org/abs/hep-ph/9409265}{{\ttfamily hep-ph/9409265}}].

\bibitem{Britto:2004nc}
R.~Britto, F.~Cachazo and B.~Feng, \emph{{Generalized unitarity and one-loop
  amplitudes in N=4 super-Yang-Mills}},
  \href{https://doi.org/10.1016/j.nuclphysb.2005.07.014}{\emph{Nucl. Phys. B}
  {\bfseries 725} (2005) 275}
  [\href{https://arxiv.org/abs/hep-th/0412103}{{\ttfamily hep-th/0412103}}].

\bibitem{Britto:2005ha}
R.~Britto, E.~Buchbinder, F.~Cachazo and B.~Feng, \emph{{One-loop amplitudes of
  gluons in SQCD}},
  \href{https://doi.org/10.1103/PhysRevD.72.065012}{\emph{Phys. Rev. D}
  {\bfseries 72} (2005) 065012}
  [\href{https://arxiv.org/abs/hep-ph/0503132}{{\ttfamily hep-ph/0503132}}].

\bibitem{Ossola:2006us}
G.~Ossola, C.G.~Papadopoulos and R.~Pittau, \emph{{Reducing full one-loop
  amplitudes to scalar integrals at the integrand level}},
  \href{https://doi.org/10.1016/j.nuclphysb.2006.11.012}{\emph{Nucl. Phys. B}
  {\bfseries 763} (2007) 147}
  [\href{https://arxiv.org/abs/hep-ph/0609007}{{\ttfamily hep-ph/0609007}}].

\bibitem{Gluza:2010ws}
J.~Gluza, K.~Kajda and D.A.~Kosower, \emph{{Towards a Basis for Planar Two-Loop
  Integrals}}, \href{https://doi.org/10.1103/PhysRevD.83.045012}{\emph{Phys.
  Rev. D} {\bfseries 83} (2011) 045012}
  [\href{https://arxiv.org/abs/1009.0472}{{\ttfamily 1009.0472}}].

\bibitem{Peraro:2019svx}
T.~Peraro, \emph{{FiniteFlow: multivariate functional reconstruction using
  finite fields and dataflow graphs}},
  \href{https://doi.org/10.1007/JHEP07(2019)031}{\emph{JHEP} {\bfseries 07}
  (2019) 031} [\href{https://arxiv.org/abs/1905.08019}{{\ttfamily
  1905.08019}}].

\bibitem{Chestnov:2022alh}
V.~Chestnov, F.~Gasparotto, M.K.~Mandal, P.~Mastrolia, S.J.~Matsubara-Heo,
  H.J.~Munch et~al., \emph{{Macaulay Matrix for Feynman Integrals: Linear
  Relations and Intersection Numbers}},
  \href{https://arxiv.org/abs/2204.12983}{{\ttfamily 2204.12983}}.

\bibitem{Mastrolia:2018uzb}
P.~Mastrolia and S.~Mizera, \emph{{Feynman Integrals and Intersection Theory}},
  \href{https://doi.org/10.1007/JHEP02(2019)139}{\emph{JHEP} {\bfseries 02}
  (2019) 139} [\href{https://arxiv.org/abs/1810.03818}{{\ttfamily
  1810.03818}}].

\bibitem{Frellesvig:2019kgj}
H.~Frellesvig, F.~Gasparotto, S.~Laporta, M.K.~Mandal, P.~Mastrolia,
  L.~Mattiazzi et~al., \emph{{Decomposition of Feynman Integrals on the Maximal
  Cut by Intersection Numbers}},
  \href{https://doi.org/10.1007/JHEP05(2019)153}{\emph{JHEP} {\bfseries 05}
  (2019) 153} [\href{https://arxiv.org/abs/1901.11510}{{\ttfamily
  1901.11510}}].

\bibitem{Frellesvig:2019uqt}
H.~Frellesvig, F.~Gasparotto, M.K.~Mandal, P.~Mastrolia, L.~Mattiazzi and
  S.~Mizera, \emph{{Vector Space of Feynman Integrals and Multivariate
  Intersection Numbers}},
  \href{https://doi.org/10.1103/PhysRevLett.123.201602}{\emph{Phys. Rev. Lett.}
  {\bfseries 123} (2019) 201602}
  [\href{https://arxiv.org/abs/1907.02000}{{\ttfamily 1907.02000}}].

\bibitem{Weinzierl:2020xyy}
S.~Weinzierl, \emph{{On the computation of intersection numbers for twisted
  cocycles}}, \href{https://doi.org/10.1063/5.0054292}{\emph{J. Math. Phys.}
  {\bfseries 62} (2021) 072301}
  [\href{https://arxiv.org/abs/2002.01930}{{\ttfamily 2002.01930}}].

\bibitem{Mizera:2019ose}
S.~Mizera, \emph{{Status of Intersection Theory and Feynman Integrals}},
  \href{https://doi.org/10.22323/1.383.0016}{\emph{PoS} {\bfseries MA2019}
  (2019) 016} [\href{https://arxiv.org/abs/2002.10476}{{\ttfamily
  2002.10476}}].

\bibitem{Frellesvig:2020qot}
H.~Frellesvig, F.~Gasparotto, S.~Laporta, M.K.~Mandal, P.~Mastrolia,
  L.~Mattiazzi et~al., \emph{{Decomposition of Feynman Integrals by
  Multivariate Intersection Numbers}},
  \href{https://doi.org/10.1007/JHEP03(2021)027}{\emph{JHEP} {\bfseries 03}
  (2021) 027} [\href{https://arxiv.org/abs/2008.04823}{{\ttfamily
  2008.04823}}].

\bibitem{Liu:2018dmc}
X.~Liu and Y.-Q.~Ma, \emph{{Determining arbitrary Feynman integrals by vacuum
  integrals}}, \href{https://doi.org/10.1103/PhysRevD.99.071501}{\emph{Phys.
  Rev. D} {\bfseries 99} (2019) 071501}
  [\href{https://arxiv.org/abs/1801.10523}{{\ttfamily 1801.10523}}].

\bibitem{Guan:2019bcx}
X.~Guan, X.~Liu and Y.-Q.~Ma, \emph{{Complete reduction of integrals in
  two-loop five-light-parton scattering amplitudes}},
  \href{https://doi.org/10.1088/1674-1137/44/9/093106}{\emph{Chin. Phys. C}
  {\bfseries 44} (2020) 093106}
  [\href{https://arxiv.org/abs/1912.09294}{{\ttfamily 1912.09294}}].

\bibitem{Larsen:2015ped}
K.J.~Larsen and Y.~Zhang, \emph{{Integration-by-parts reductions from unitarity
  cuts and algebraic geometry}},
  \href{https://doi.org/10.1103/PhysRevD.93.041701}{\emph{Phys. Rev. D}
  {\bfseries 93} (2016) 041701}
  [\href{https://arxiv.org/abs/1511.01071}{{\ttfamily 1511.01071}}].

\bibitem{Larsen:2016tdk}
K.J.~Larsen and Y.~Zhang, \emph{{Integration-by-parts reductions from the
  viewpoint of computational algebraic geometry}},
  \href{https://doi.org/10.22323/1.260.0029}{\emph{PoS} {\bfseries LL2016}
  (2016) 029} [\href{https://arxiv.org/abs/1606.09447}{{\ttfamily
  1606.09447}}].

\bibitem{Zhang:2016kfo}
Y.~Zhang, \emph{{Lecture Notes on Multi-loop Integral Reduction and Applied
  Algebraic Geometry}},  12, 2016
  [\href{https://arxiv.org/abs/1612.02249}{{\ttfamily 1612.02249}}].

\bibitem{Georgoudis:2016wff}
A.~Georgoudis, K.J.~Larsen and Y.~Zhang, \emph{{Azurite: An algebraic geometry
  based package for finding bases of loop integrals}},
  \href{https://doi.org/10.1016/j.cpc.2017.08.013}{\emph{Comput. Phys. Commun.}
  {\bfseries 221} (2017) 203}
  [\href{https://arxiv.org/abs/1612.04252}{{\ttfamily 1612.04252}}].

\bibitem{Georgoudis:2017iza}
A.~Georgoudis, K.J.~Larsen and Y.~Zhang, \emph{{Cristal and Azurite: new tools
  for integration-by-parts reductions}},
  \href{https://doi.org/10.22323/1.290.0020}{\emph{PoS} {\bfseries RADCOR2017}
  (2017) 020} [\href{https://arxiv.org/abs/1712.07510}{{\ttfamily
  1712.07510}}].

\bibitem{Bohm:2017qme}
J.~B\"ohm, A.~Georgoudis, K.J.~Larsen, M.~Schulze and Y.~Zhang, \emph{{Complete
  sets of logarithmic vector fields for integration-by-parts identities of
  Feynman integrals}},
  \href{https://doi.org/10.1103/PhysRevD.98.025023}{\emph{Phys. Rev. D}
  {\bfseries 98} (2018) 025023}
  [\href{https://arxiv.org/abs/1712.09737}{{\ttfamily 1712.09737}}].

\bibitem{Bohm:2018bdy}
J.~B\"ohm, A.~Georgoudis, K.J.~Larsen, H.~Sch\"onemann and Y.~Zhang,
  \emph{{Complete integration-by-parts reductions of the non-planar hexagon-box
  via module intersections}},
  \href{https://doi.org/10.1007/JHEP09(2018)024}{\emph{JHEP} {\bfseries 09}
  (2018) 024} [\href{https://arxiv.org/abs/1805.01873}{{\ttfamily
  1805.01873}}].

\bibitem{Bendle:2019csk}
D.~Bendle, J.~B\"ohm, W.~Decker, A.~Georgoudis, F.-J.~Pfreundt, M.~Rahn et~al.,
  \emph{{Integration-by-parts reductions of Feynman integrals using Singular
  and GPI-Space}}, \href{https://doi.org/10.1007/JHEP02(2020)079}{\emph{JHEP}
  {\bfseries 02} (2020) 079}
  [\href{https://arxiv.org/abs/1908.04301}{{\ttfamily 1908.04301}}].

\bibitem{Boehm:2020zig}
J.~Boehm, D.~Bendle, W.~Decker, A.~Georgoudis, F.-J.~Pfreundt, M.~Rahn et~al.,
  \emph{{Module Intersection for the Integration-by-Parts Reduction of
  Multi-Loop Feynman Integrals}},
  \href{https://doi.org/10.22323/1.383.0004}{\emph{PoS} {\bfseries MA2019}
  (2022) 004} [\href{https://arxiv.org/abs/2010.06895}{{\ttfamily
  2010.06895}}].

\bibitem{Bendle:2021ueg}
D.~Bendle, J.~Boehm, M.~Heymann, R.~Ma, M.~Rahn, L.~Ristau et~al.,
  \emph{{Two-loop five-point integration-by-parts relations in a usable form}},
   \href{https://arxiv.org/abs/2104.06866}{{\ttfamily 2104.06866}}.

\bibitem{Feng:2022iuc}
B.~Feng and T.~Li, \emph{{PV-Reduction of Sunset Topology with Auxiliary
  Vector}},  \href{https://arxiv.org/abs/2203.16881}{{\ttfamily 2203.16881}}.

\bibitem{Schabinger:2011dz}
R.M.~Schabinger, \emph{{A New Algorithm For The Generation Of
  Unitarity-Compatible Integration By Parts Relations}},
  \href{https://doi.org/10.1007/JHEP01(2012)077}{\emph{JHEP} {\bfseries 01}
  (2012) 077} [\href{https://arxiv.org/abs/1111.4220}{{\ttfamily 1111.4220}}].

\bibitem{Baikov:1996iu}
P.A.~Baikov, \emph{{Explicit solutions of the multiloop integral recurrence
  relations and its application}},
  \href{https://doi.org/10.1016/S0168-9002(97)00126-5}{\emph{Nucl. Instrum.
  Meth. A} {\bfseries 389} (1997) 347}
  [\href{https://arxiv.org/abs/hep-ph/9611449}{{\ttfamily hep-ph/9611449}}].

\bibitem{DGPS}
W.~Decker, G.-M.~Greuel, G.~Pfister and H.~Sch\"onemann, ``{\sc Singular}
  {4-3-0} --- {A} computer algebra system for polynomial computations.''
  \url{http://www.singular.uni-kl.de}, 2022.

\bibitem{Feng:2021enk}
B.~Feng, T.~Li and X.~Li, \emph{{Analytic tadpole coefficients of one-loop
  integrals}}, \href{https://doi.org/10.1007/JHEP09(2021)081}{\emph{JHEP}
  {\bfseries 09} (2021) 081}
  [\href{https://arxiv.org/abs/2107.03744}{{\ttfamily 2107.03744}}].

\bibitem{Hu:2021nia}
C.~Hu, T.~Li and X.~Li, \emph{{One-loop Feynman integral reduction by
  differential operators}},
  \href{https://doi.org/10.1103/PhysRevD.104.116014}{\emph{Phys. Rev. D}
  {\bfseries 104} (2021) 116014}
  [\href{https://arxiv.org/abs/2108.00772}{{\ttfamily 2108.00772}}].

\bibitem{Feng:2022rwj}
B.~Feng, J.~Gong and T.~Li, \emph{{Universal Treatment of Reduction for
  One-Loop Integrals in Projective Space}},
  \href{https://arxiv.org/abs/2204.03190}{{\ttfamily 2204.03190}}.

\bibitem{Feng:2022rfz}
B.~Feng, C.~Hu, T.~Li and Y.~Song, \emph{{Reduction with Degenerate Gram matrix
  for One-loop Integrals}},  \href{https://arxiv.org/abs/2205.03000}{{\ttfamily
  2205.03000}}.

\bibitem{Lee:2013hzt}
R.N.~Lee and A.A.~Pomeransky, \emph{{Critical points and number of master
  integrals}}, \href{https://doi.org/10.1007/JHEP11(2013)165}{\emph{JHEP}
  {\bfseries 11} (2013) 165} [\href{https://arxiv.org/abs/1308.6676}{{\ttfamily
  1308.6676}}].

\bibitem{Chen:2022lzr}
J.~Chen, X.~Jiang, C.~Ma, X.~Xu and L.L.~Yang, \emph{{Baikov representations,
  intersection theory, and canonical Feynman integrals}},
  \href{https://arxiv.org/abs/2202.08127}{{\ttfamily 2202.08127}}.

\end{thebibliography}\endgroup

\end{document}